\documentclass[a4paper,english,conference]{IEEEtran}
\usepackage{latexsym}
\usepackage{float}
\usepackage{amsfonts}
\usepackage{amsbsy}
\usepackage{amssymb}
\usepackage{times}
\usepackage{graphicx}
\usepackage{enumerate}
\usepackage[usenames]{color}
\usepackage[dvips]{pstcol}
\usepackage{epstopdf}
\usepackage{cite}
\usepackage{amssymb}
\usepackage{amsfonts}
\usepackage{graphicx}
\usepackage{epsfig}
\usepackage{psfrag}
\usepackage{xcolor}
\usepackage{amsfonts, bm}
\usepackage{epstopdf}
\usepackage{cite}
\usepackage{color}
\usepackage{xcolor}
\usepackage{subfig}
\usepackage{verbatim}
\usepackage{multirow}
\usepackage{array}
\usepackage{booktabs}
\usepackage{amsthm}
\usepackage{makecell}

\usepackage[linesnumbered, ruled]{algorithm2e}
\usepackage{algpseudocode}
\usepackage{amsmath}

\IEEEoverridecommandlockouts
\begin{document}
	\title{ A Unified Multi-Task Semantic Communication System with Domain Adaptation}
	\author{\IEEEauthorblockN{ Guangyi Zhang \IEEEauthorrefmark{1}, Qiyu Hu \IEEEauthorrefmark{1}, Zhijin Qin \IEEEauthorrefmark{2}, Yunlong Cai \IEEEauthorrefmark{1}, and Guanding Yu \IEEEauthorrefmark{1}  }
		\IEEEauthorblockA{\IEEEauthorrefmark{1} College of Information Science and Electronic Engineering, Zhejiang University, Hangzhou, China }
		\IEEEauthorblockA{\IEEEauthorrefmark{2} School of Electronic Engineering and Computer Science, Queen Mary University of London, London, UK  \\ E-mail: \{zhangguangyi, qiyhu, ylcai, yuguanding\}@zju.edu.cn, z.qin@qmul.ac.uk} }
	
	\maketitle
	\vspace{-3.3em}
	\begin{abstract}
	The task-oriented semantic communication systems have achieved significant performance gain, however, the paradigm that employs a model for a specific task might be limited, since the system has to be updated once the task is changed or multiple models are stored for serving various tasks. To address this issue, we firstly propose a unified deep learning enabled semantic communication system (U-DeepSC), where a unified model is developed to serve various transmission tasks. To jointly serve these tasks in one model with fixed parameters, we employ domain adaptation in the training procedure to specify the task-specific features for each task. Thus, the system only needs to transmit the task-specific features, rather than all the features, to reduce the transmission overhead. Moreover, since each task is of different difficulty and requires different number of layers to achieve satisfactory performance, we develop the multi-exit architecture to provide early-exit results for relatively simple tasks. In the experiments, we employ a proposed U-DeepSC to serve five tasks with multi-modalities. Simulation results demonstrate that our proposed U-DeepSC achieves comparable performance to the task-oriented semantic communication system designed for a specific task with significant transmission overhead reduction and much less number of model parameters.
	\end{abstract}
	\IEEEpeerreviewmaketitle
	
	\section{Introduction}
	
	With the development of wireless communication and machine learning, a huge amount of intelligent applications have appeared in the networks \cite{IoTData}. To support massive connectivity for these applications over limited wireless resources, the conventional communication systems are facing critical challenges. To address this issue, semantic communications have been considered as a promising technology to achieve better performance\cite{Principle, SemanMaga}.  Different from conventional communications, semantic communications only take into account the relevant semantic features to the tasks, which enables the systems to recover information from the received semantic features. 
	
	According to the task types at the receiver, the existing works on semantic communications can be mainly divided into two categories: data reconstruction \cite{JSCC,DeepSC, device,wit} and task execution \cite{ImagRetri, TransIoT, vqvae, Task-oriented,  JSCCf}. As for the data reconstruction, the semantic system extracts global semantic information of source data. Specifically, The authors in \cite{JSCC} proposed a joint source channel coding (JSCC) system for image transmission. For text transmission, a so called DeepSC framework, has been proposed to encode the text information into various length by employing sentence information \cite{DeepSC}. In \cite{wit},  an attention based JSCC has been proposed to operate with different signal-to-noise (SNR) levels during image transmission. 
	
	For the task execution applications, only the task-specific semantic information is extracted and encoded at the transmitter \cite{ImagRetri, TransIoT, vqvae,  Task-oriented,JSCCf}. In particular, the authors of \cite{ImagRetri} proposed a model for image retrieval task under power and bandwidth constraints. In \cite{TransIoT}, an image classification-oriented semantic communication system has been developed. The authors of \cite{vqvae} have proposed a  vector quantization-variational autoencoder (VQ-VAE) based robust semantic communication systems for image classification. 
	\begin{figure*}[!htbp]
		\begin{centering}
			\includegraphics[width=0.96\textwidth]{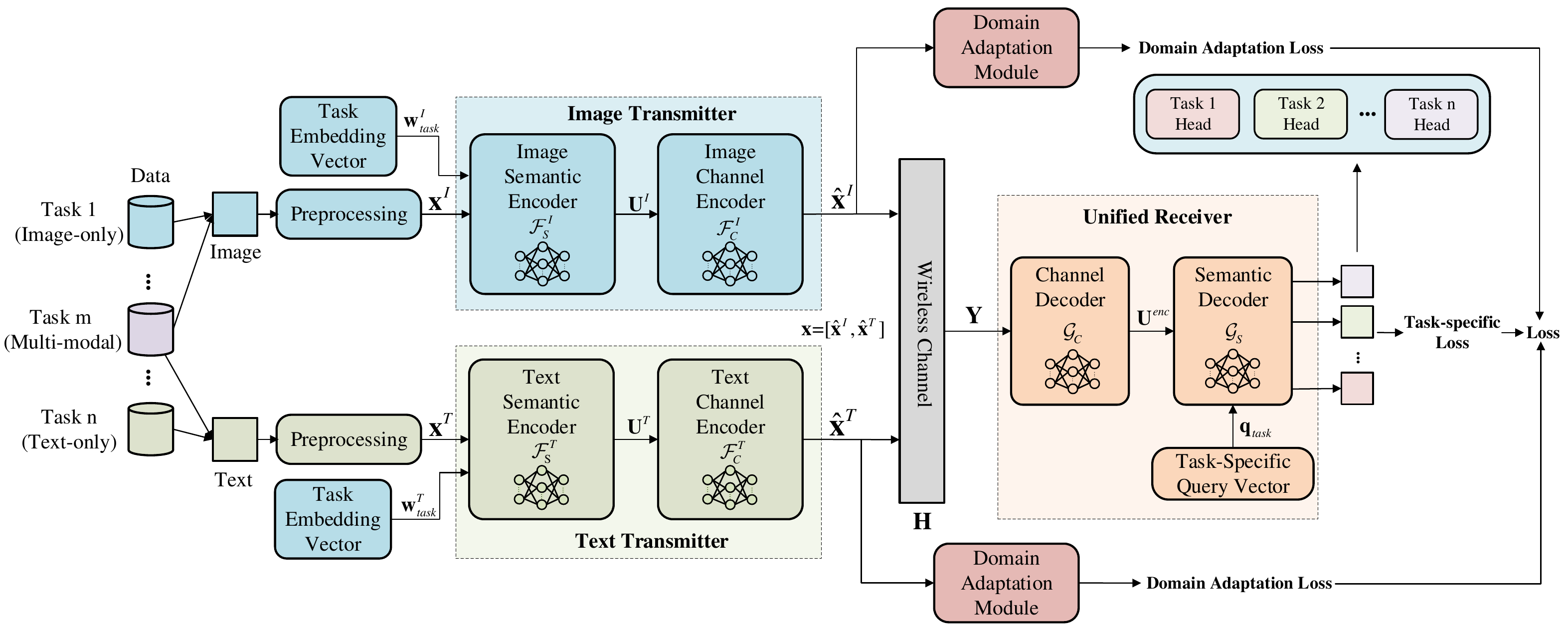}
			\par\end{centering}
		\caption{The framework of the proposed unified deep learning enabled semantic communication system.}
		\label{FrameArchitect}
	\end{figure*}
	 Compared with the data reconstruction applications, the transmission overhead can be further reduced in the task execution applications. However, they can only handle one task with single modality of data. Therefore, it is difficult for serving various tasks in practice for two reasons: (i) The model in the system has to be updated once the task is changed, which leads to a lot of gradient transmission for retraining the model, since the model at the transmitter and receiver are jointly trained; (ii) Multiple models are stored for serving different tasks, which is difficult for the devices with limited storage resources. In \cite{Task-oriented}, a Transformer-based framework has been proposed to address this issue initially. It is able to share the same transmitter structures for the considered tasks. However, the model in \cite{Task-oriented} still needs to be retrained separately for different tasks, and the architecture of the receiver hasn’t been unified for different tasks yet. Inspired from multi-task learning \cite{Unit}, we firstly propose a unified deep learning enabled semantic communication system (U-DeepSC) to further address this issue by unifying the transmitter and receiver simultaneously.
	 
	 To the best of our knowledge, this is the first work on the unified semantic communication system for serving various tasks. In this paper, we propose the U-DeepSC, an encoder-decoder semantic communication architecture, to serve multiple tasks with different modalities. Our proposed model is able to simultaneously deal with a number of tasks consisting of image-only and text-only tasks, and even image-and-text reasoning tasks with two modalities of data. In order to extract and transmit only the task-specific information in the U-DeepSC, we divide the encoded features into different parts according to the tasks, and each part corresponds to the semantic information of one specific task. To further specify the semantic information, we employ the domain adaptation \cite{domain_survery} to project the features of different tasks into specific feature domains. In particular, we propose the domain adaptation loss for the joint training procedure. Moreover, since each task is of different difficulty and requires different number of layers to achieve satisfactory performance, the multi-exit architecture is developed by inserting the early-exit modules after the intermediate layer of the decoder to provide early-exit results for relatively simple tasks \cite{MultiExit}.  Simulation results show that our proposed method achieves comparable performance to the task-oriented semantic communication systems designed for a specific task with much reduced transmission overhead and fewer model parameters.

	The rest of this paper is structured as follows. Section II introduces the framework of U-DeepSC. In Section III, the detailed architecture of the proposed U-DeepSC is presented. Simulation results are presented in Section IV. Finally, Section V concludes this paper.



\section{Framework of U-DeepSC} \label{System}

In this section, we propose the framework of U-DeepSC, and the details of the considered tasks.
\subsection{System Model}
As shown in Fig. \ref{FrameArchitect}, the proposed U-DeepSC is able to handle a number of tasks with two modalities, i.e., image and text. The proposed framework mainly consists of three parts: image transmitter, text transmitter, and unified receiver. The deep neural networks (DNNs) are employed to represent the transmitter and the unified receiver. In particular, the image transmitter consists of the image semantic encoder and the image channel encoder, while the text transmitter consists of the text semantic encoder. Moreover, the receiver consists of the unified channel decoder and the unified semantic decoder.

We consider a communication system equipped with $N_t$ transmit antennas and $N_r$ receive antennas. The inputs of the system are image, $\mathbf{x}^{I}$, and text, $\mathbf{x}^{T}$. The image semantic encoder learns to map $\mathbf{x}^{I}$ into the encoded image features, while $\mathbf{x}^{T}$ is processed by the text semantic encoder to obtain the encoded text features. Thus, the encoded features of image and text can be represented by
\begin{equation}
	\hat{\mathbf{x}}^{I}=\mathcal{F}^{I}_{C} \big( \mathcal{F}^{I}_{S}(\mathbf{x}^{I};\bm{\theta} ^{I}_{S}); \bm{\theta} ^{I}_{C}\big),
\end{equation}
and 
\begin{equation}
	\hat{\mathbf{x}}^{T}=\mathcal{F}^{T}_{C} \big( \mathcal{F}^{T}_{S}(\mathbf{x}^{T};\bm{\theta} ^{T}_{S}); \bm{\theta} ^{I}_{C}\big),
\end{equation}
respectively, where $\hat{\mathbf{x}}^{I}\in \mathbb{C}^{N_{t}\times 1}$, $\hat{\mathbf{x}}^{T}\in \mathbb{C}^{N_{t}\times 1}$, $\bm{\theta} ^{I}_{S}$, and $\bm{\theta} ^{I}_{C}$ denote the trainable parameters of the image semantic encoder, $\mathcal{F}^{I}_{S}$, and the image channel encoder, $\mathcal{F}^{I}_{C}$, respectively, $\bm{\theta} ^{T}_{S}$ and $\bm{\theta} ^{T}_{C}$ denote the trainable parameters of the text semantic encoder, $\mathcal{F}^{T}_{S}$, and the text channel encoder, $\mathcal{F}^{T}_{C}$, respectively. We concatenate the encoded features to obtain the transmitted symbol streams expressed as $\mathbf{x} = \big[\hat{\mathbf{x}}^{I}, \hat{\mathbf{x}}^{T} \big ].$
Then, the received signal at the receiver is given by
\begin{equation}
	\mathbf{Y} = \mathbf{H} \mathbf{x} + \mathbf{n},
\end{equation}
where $\mathbf{H} \in \mathbb{C}^{N_{r}\times N_{t}}$ represents the channel gain and $\mathbf{n}\sim \mathcal{CN}(\bm{0}, \sigma^{2}\mathbf{I})$ is the additive white Gaussian noise (AWGN).

At the receiver, the decoded signal can be represented as
\begin{equation}
	\hat{\mathbf{Y}}=\mathcal{G}_{S} \big(\mathcal{G}_{C}(\mathbf{Y};\bm{\phi}_{C}); \bm{\phi}_{S}\big),
\end{equation}
where $\bm{\phi}_{C}$ and $\bm{\phi}_{S}$ denote the trainable parameters of the channel decoder, $\mathcal{G}_{C}$, and the semantic decoder, $\mathcal{G}_{S}$, respectively. Finally, the obtained features are further processed by the light-weight task-specific heads to execute downstream tasks. Particularly, the task-specific head refers to some simple layers that reshape the decoded features into the intended  dimension of output, e.g., the number of classes for a classification task.

\subsection{Task Description} 
To provide a thorough analysis of U-DeepSC and also provide sufficient results to prove the effectiveness, we will experiment with jointly handling prominent tasks from different domains, including sentiment analysis, visual question answering (VQA),  image retrieval, image data reconstruction, and text data reconstruction tasks. Besides, these tasks have been widely considered in existing semantic comunication systems \cite{DeepSC,Task-oriented,ImagRetri}.

\subsubsection{Sentiment analysis}
The purpose of the sentiment analysis task is to classify whether the sentiment of a given sentence is positive or negative. It is essentially the binary classification problem. Thus, we take classification accuracy as the performance metric for sentiment analysis and VQA, and the cross entropy as the loss function to train the model. 

\subsubsection{VQA}
In VQA task, the images and questions in text are processed by the model to classify which answer is right. Thus, we take answer accuracy as the performance metric and the cross entropy as the loss function. 

\subsubsection{Image retrieval}
The image retrieval task aims at finding similar images to a query image among the images stored in a large server. To evaluate the performance of image retrieval task, the Recall@1 is adopted as the performance evaluation metric, which refers to the ratio of successful image retrieval at the first query. To learn this task, we adopt the triplet loss, which is given by 
\begin{equation}
	\mathcal{L}=\max (d(\mathbf{s}_a,\mathbf{s}_p),d(\mathbf{s}_a,\mathbf{s}_n)+m,0),
\end{equation}
where  $m$ is a constant, $\mathbf{s}_p$ denotes the positive sample with the same class as sample $\mathbf{s}_a$, $\mathbf{s}_n$ is the negative sample with the different class from $\mathbf{s}_a$, and $d$ is the distance metric. This triplet loss aims to make the distance between the features of two similar samples closer, and to make the distance between the features of two different samples farther. Thus, the model will be able to find the similar samples to the given input according to their encoded features.
\subsubsection{Image reconstruction}
The performance of the image reconstruction task is quantified by the peak signal-to-noise ratio (PSNR). The PSNR measures the ratio between the maximum possible power and the noise, which is given by
\begin{equation}
	\textrm{PSNR}=10 \log_{10}{\frac{\textrm{MAX}^{2}}{\textrm{MSE}}}(\textrm{dB}),
\end{equation}
where $\textrm{MSE}=d(\mathbf{x},\hat{\mathbf{x}})$ denotes the mean squared-error (MSE) between the source image, $\mathbf{x}$, and the reconstructed image, $\hat{\mathbf{x}}$, and $\textrm{MAX}$ is the maximum possible value of the pixels. Moreover, the MSE is adopted as the training loss.

\subsubsection{Text reconstruction}
As for the text reconstruction task, the bi-lingual evaluation understudy (BLEU) score is adopted to measure the performance. BLEU score is a scalar between $0$ and $1$, which evaluates the similarity between the reconstructed text and the source text, with $1$ representing highest similarity. We take the cross entropy as the loss function since the BLEU score is non-differentiable.

\section{Architecture of the Proposed U-DeepSC}  \label{Architect}
In this section, we design the architecture of U-DeepSC. The U-DeepSC is built based on the unified Transformer structure \cite{Unit}, which consists of the separate semantic/channel encoders for each modality and the unified semantic/channel decoder with light-weight task-specific heads and multi-exit module.

\subsection{Semantic Encoder} 
Since the data from different modalities have totally different statistical characteristics, semantic information, and encoded features, we design an image semantic encoder and a text semantic encoder for image and text, respectively.
\subsubsection{Image semantic encoder}
The image-only and multi-modal tasks take an image $\mathbf{I}$ as input, and the extracted features $\mathbf{x}^{I}$ is given by $\mathbf{x}^{I} = f(\mathbf{I})$, where  $f$ denotes the preprocessing module. Then, a Transformer encoder is employed as the image semantic encoder to encode $\mathbf{x}^{I}$ to encoded image feature matrix, $\mathbf{U}^{I}$. Moreover, since different tasks may require the semantic encoder to extract different features, a task embedding vector, $\mathbf{w}_{task}^{I}$, is added to the semantic encoder given as $[\mathbf{x}^{I}, \mathbf{w}_{task}^{I}]$, to indicate the model which task to perform with the given sample, $\mathbf{I}$, and to allow it to extract task-specific information. Thus we obtain the encoded images feature matrix $\mathbf{U}^{I} = \{\mathbf{u}_{1}^{I},...,\mathbf{u}_{L}^{I}\}$,  where $\mathbf{u}^{I}$ denotes the encoded image feature vectors, and $L$ is the number of image feature vectors.

\subsubsection{Text semantic encoder}
As for text input, we preprocess the input text into a sequence of $S$ features $\mathbf{x}^{T}=\{ \mathbf{w}_{1}^{T},...,\mathbf{w}_{S}^{T} \}$. Subsequently, $\mathbf{x}^{T}$ is encoded by the text semantic encoder, which consists of a Transformer encoder. Similar to image semantic encoder, we also add a learned task embedding vector $\mathbf{w}_{task}^{T}$ by concatenating it at the beginning of $\mathbf{x}^{T}$. Then, the concatenated sequence, $[\mathbf{x}^{T}$, $\mathbf{w}_{task}^{T} ]$, is input into the text semantic encoder, and it outputs the encoded text features as $\mathbf{U}^{T} = \{ \mathbf{u}_{1}^{T},...,\mathbf{u}_{S}^{T}\} $.

\begin{figure*}[!htbp]
	\begin{centering}
		\includegraphics[width=0.82\textwidth]{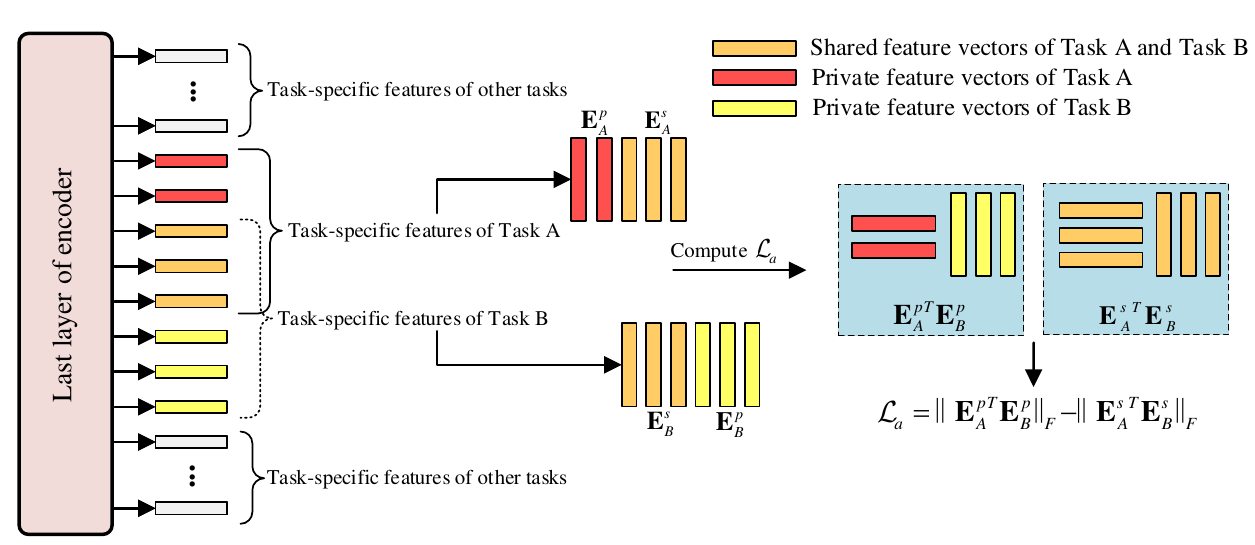}
		\par\end{centering}
	\caption{The details of domain adaptation module.}
	\label{dodo}
\end{figure*}
\begin{figure}[!h]
	\begin{centering}
	\includegraphics[width=0.42\textwidth]{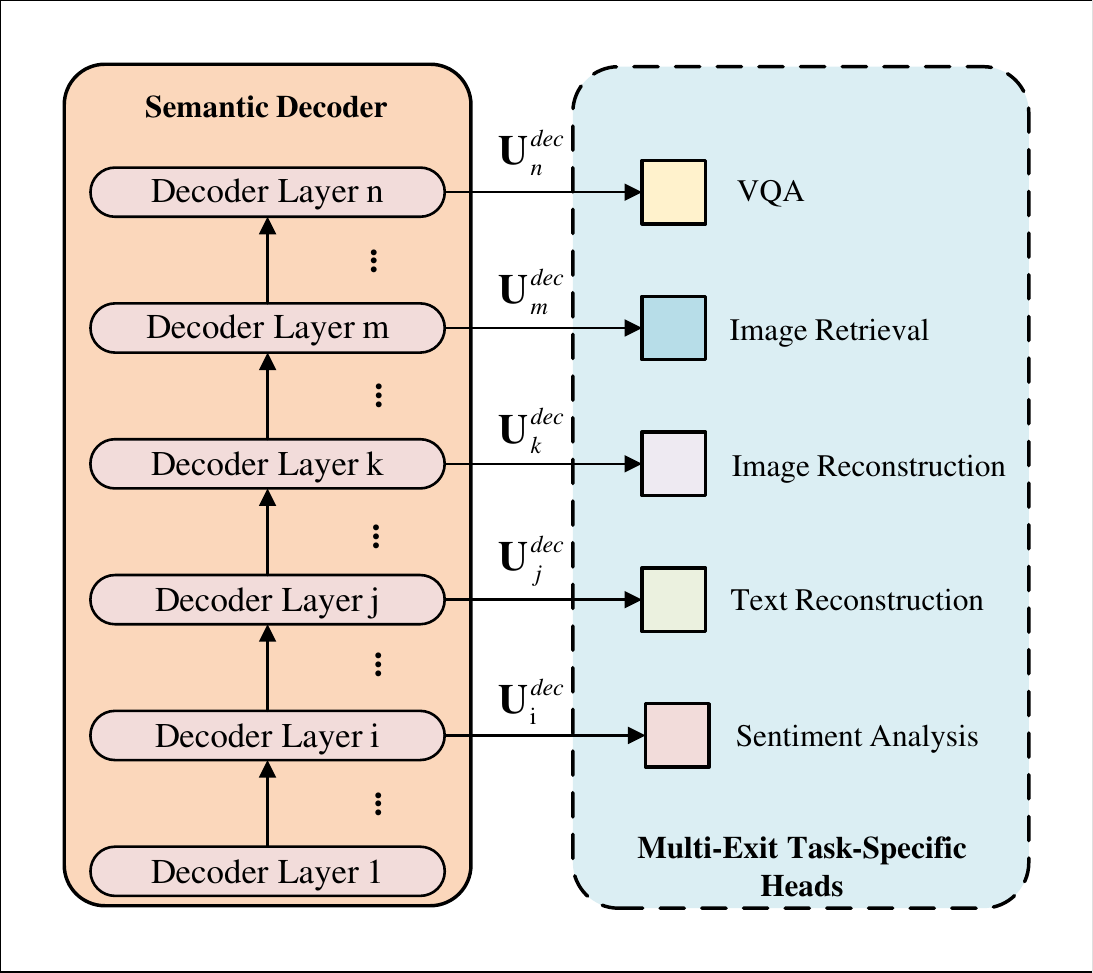}
		\par\end{centering}
\caption{The proposed structure of semantic decoder with multi-exit architecture.}
\label{Multi-exit}
\end{figure}

\subsection{Domain Adaptation Module}
Note that the overall encoded features contain the global semantic information, and they will be utilized for the aforementioned text and image reconstruction tasks. However, the other tasks, e.g., VQA and sentiment analysis, only require the task-specific semantic information. Thus, we need to specify these task-specific features for different tasks and transmit them, rather than send the overall encoded features. In order to specify the task-specific semantic information in the U-DeepSC, we divide the encoded features into several parts for different tasks as shown in the left of Fig. \ref{dodo}. Moreover, since the semantic information of different tasks may overlap, they may share part of the encoded features. Thus, the encoded features of each task can be further divided into private features and shared features. In order to enable the unified decoder to better distinguish the features of different tasks, the domain adaptation module is introduced in the training procedure to make the shared features of different tasks similar to each other.

 As shown in Fig. \ref{dodo}, we denote the shared feature matrix of task A and task B as $\mathbf{E}^{s}_{A}$ and $\mathbf{E}^{s}_{B}$, respectively, which consists of the shared encoded features. Similarly, the private feature matrix of task A and task B can denoted as $\mathbf{E}^{p}_{A}$ and $\mathbf{E}^{p}_{B}$, respectively. We define the similarity of the features as the Frobenius norm of the product of corresponding feature matrices, e.g., the similarity between the shared features can be denoted as $\Vert {\mathbf{E}^{s}_{A}}^{T}\mathbf{E}^{s}_{B} \Vert_F$, with the larger value indicating the higher similarity. Then, the adaptation loss is designed to increase the similarity between the shared features of task A and task B, as well as decreasing the similarity between private features for each tasks,  which is given by
\begin{equation} 
	\label{ada}
	\mathcal{L}_{a}=\Vert {\mathbf{E}^{p}_{A}}^{T}\mathbf{E}^{p}_{B} \Vert_F -\Vert {\mathbf{E}^{s}_{A}}^{T}\mathbf{E}^{s}_{B} \Vert_F.
\end{equation} 
 It enables the model to project the private features of different tasks into different domains, while the shared features are projected into the same domain. Thus, the task-specific semantic information can be better segmented and aggregated on the task-specific features. In this way, different tasks require different encoded features, where the model can be trained more easily and the  performance of these tasks can be improved. It also significantly reduces the transmission overhead, since only the task-specific encoded features are transmitted. For example, as shown in Fig. 2, Task A requires the red and orange features, while Task B requires the orange and yellow features. In addition, by training two tasks at one time and mapping the features to different domains, the network can avoid forgetting the past learned tasks, which is beneficial to learn multiple tasks.

\subsection{Unified Semantic Decoder with Multi-Exit Task Heads}
\subsubsection{Unified semantic decoder}
\begin{figure*}[t]
	
	\begin{centering}
		\subfloat[Sentiment Analysis]{\label{fig:a}\includegraphics[width=5.2cm]{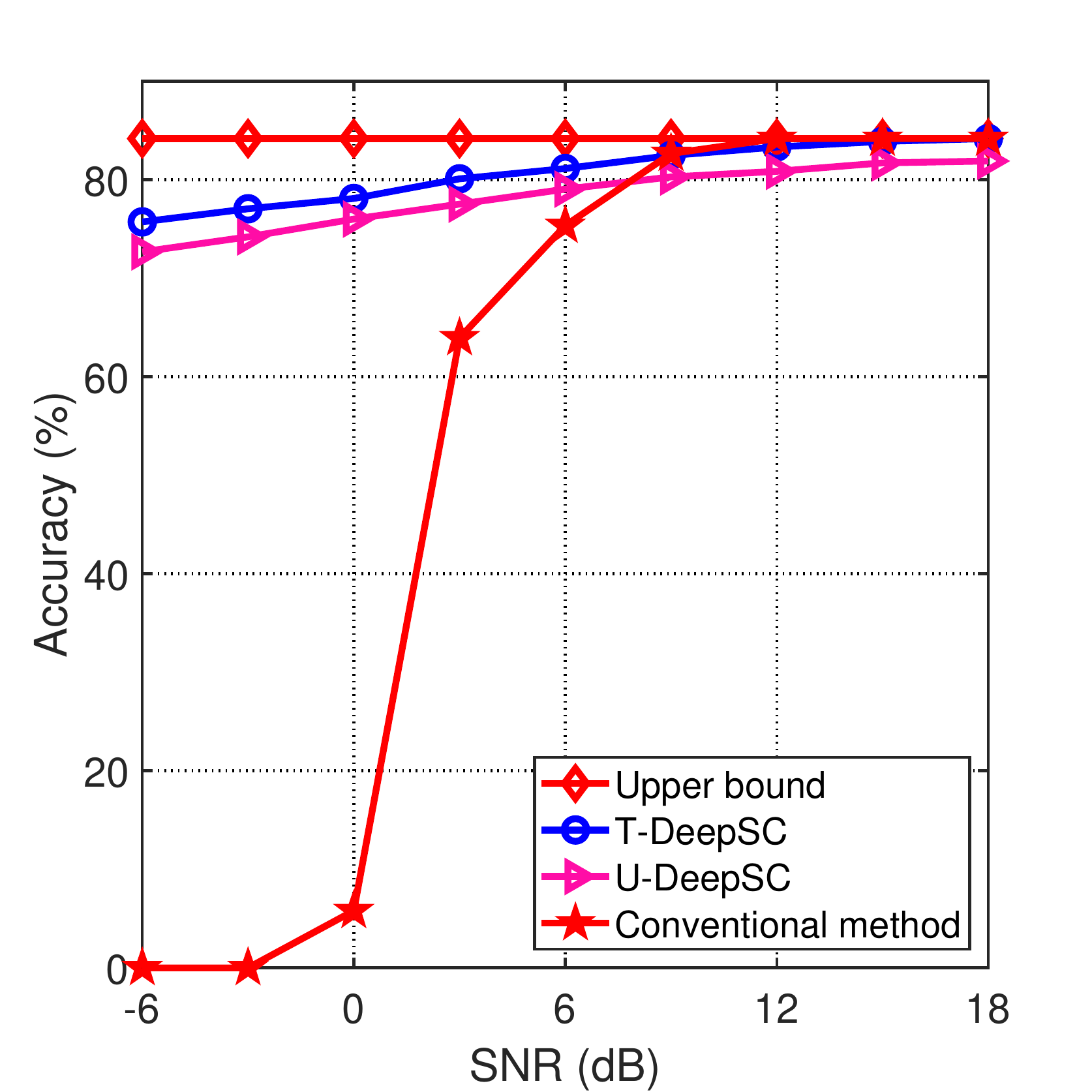}}\quad
		\subfloat[VQA]{\label{fig:b}\includegraphics[width=5.2cm]{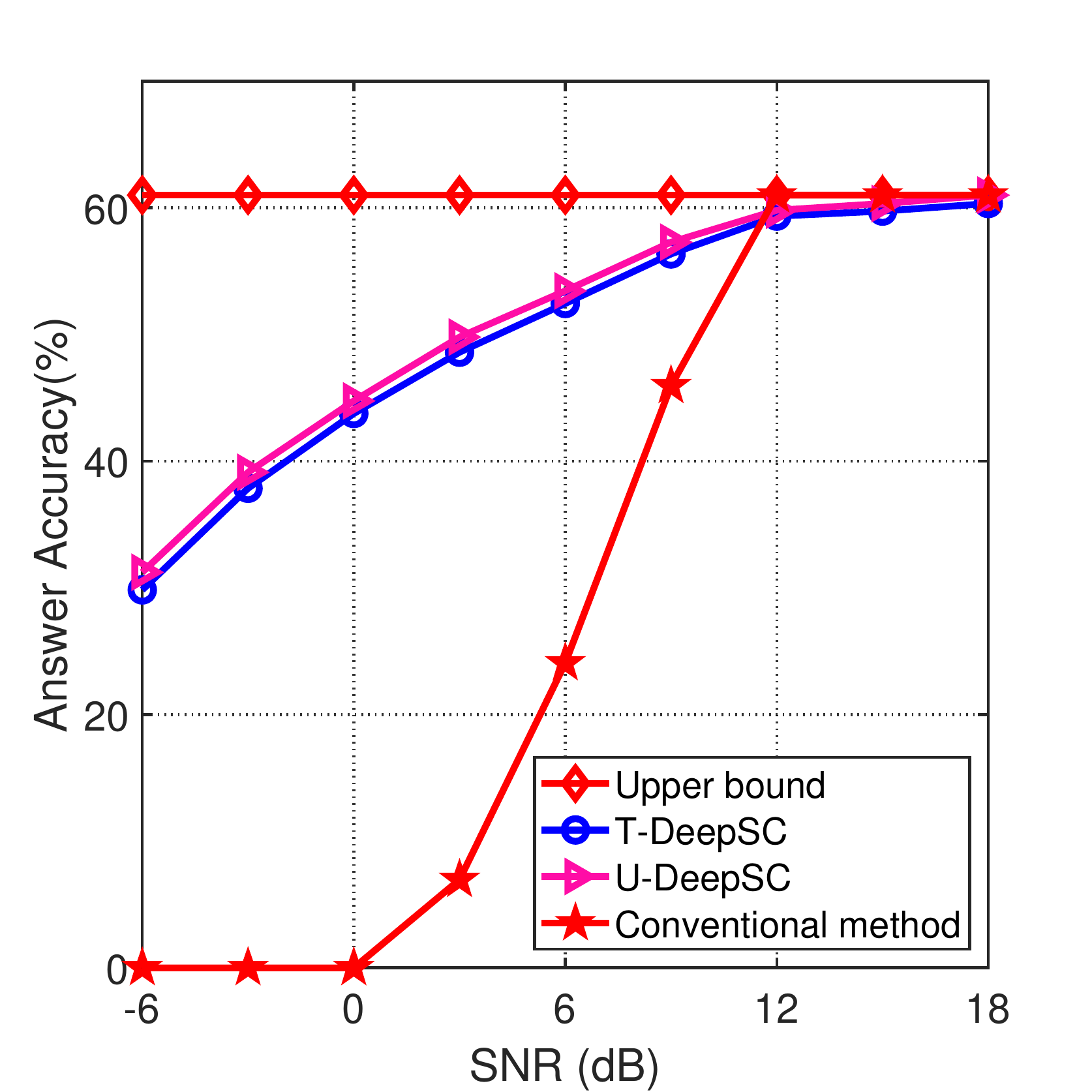}}\quad
		\subfloat[Image Retrieval]{\label{fig:a}\includegraphics[width=5.2cm]{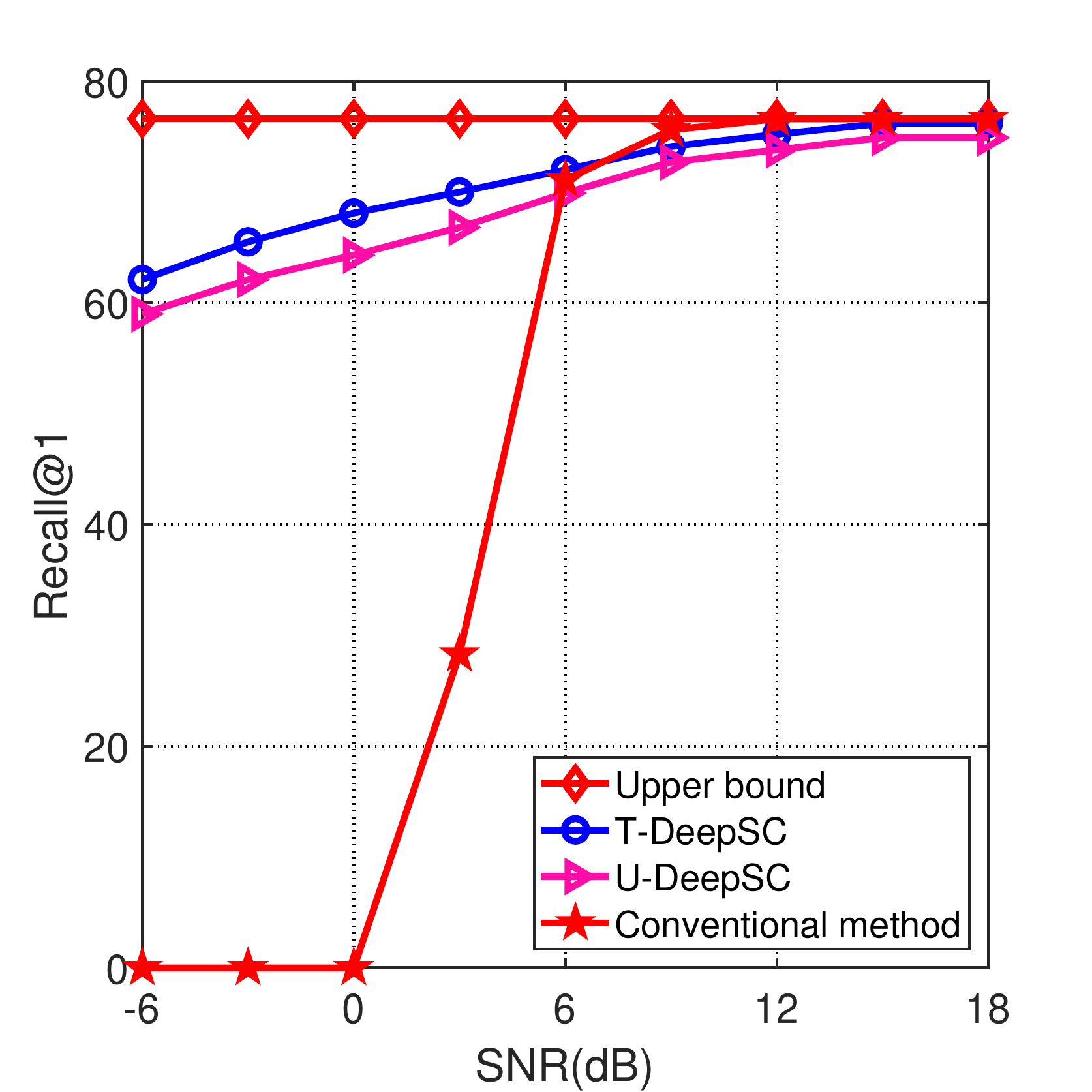}}\quad
		\subfloat[Text Reconstruction]{\label{fig:b}\includegraphics[width=5.2cm]{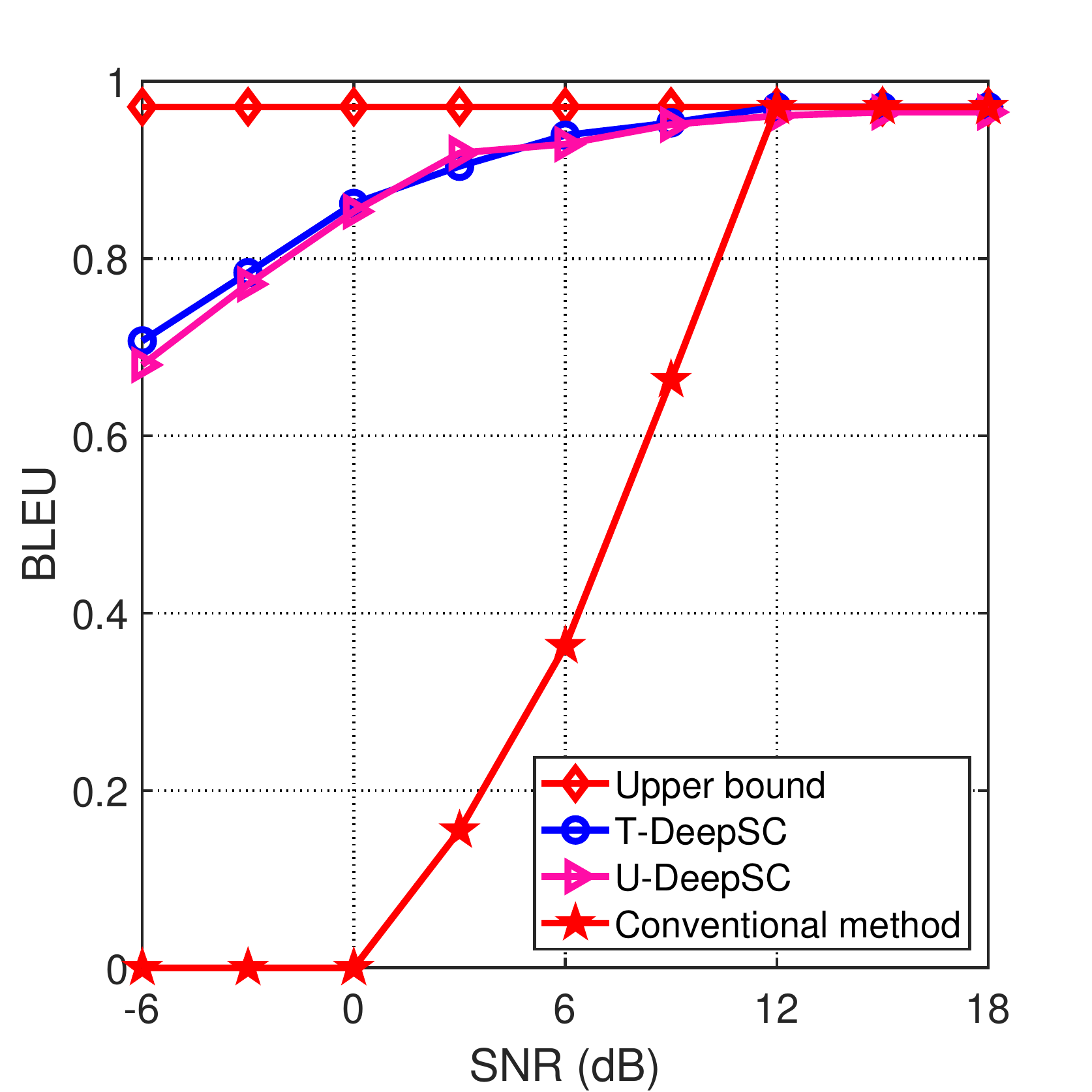}}\quad
		\subfloat[Image Reconstruction]{\label{fig:b}\includegraphics[width=5.2cm]{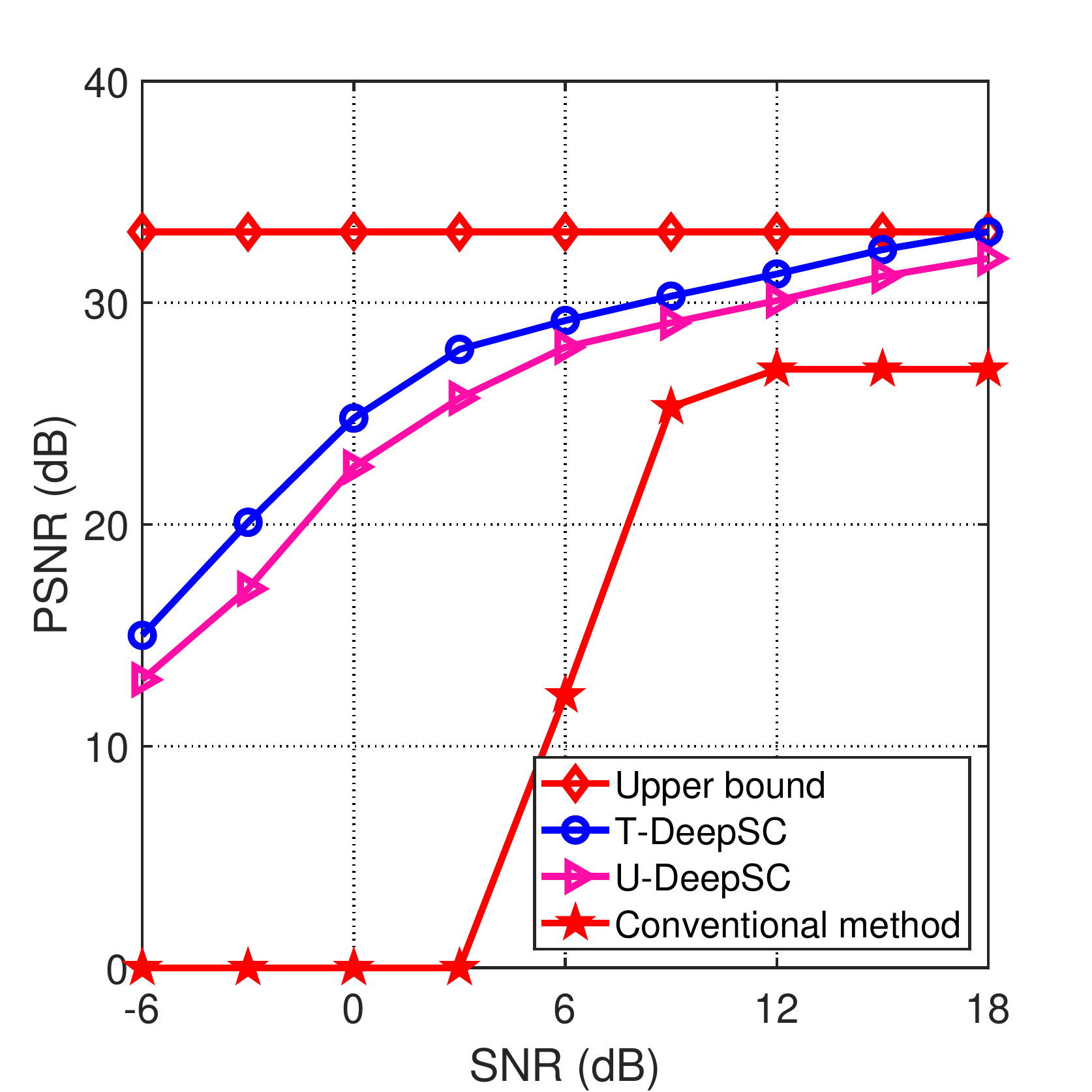}}
		
		\caption{The performance of five tasks versus SNR.} 
		\label{results}
	\end{centering}
\end{figure*}

The received encoded features are firstly processed by the channel decoder and we denote its output as $\mathbf{U}^{enc}$. For image-only tasks and text-only tasks, the input to the decoder can be represented by $\mathbf{U}^{enc}=\hat{\mathbf{U}}^{I}$ and $\mathbf{U}^{enc}=\hat{\mathbf{U}}^{T}$, where $\hat{\mathbf{U}}^{I}$ and $\hat{\mathbf{U}}^{T}$ denote the decoded image features and text features, respectively. For multi-modal tasks, we concatenate the decoded semantic features from the image and text into a sequence as $\mathbf{U}^{enc}= [\hat{\mathbf{U}}^{I}, \hat{\mathbf{U}}^{T}]$.

Unlike the separate design at the transmitter, the semantic decoder is built upon the unified Transformer decoder structure for the five tasks, as shown in Fig. \ref{FrameArchitect}. The semantic decoders take the output of channel decoder, $\mathbf{U}^{enc}$, and the task-specific query vector, $\mathbf{q}_{task}$, as input. The  task-specific query vector is able to indicate which task the semantic decoder is to handle. Then, we obtain the output of $i$-th decoder layer, $\mathbf{U}_{i}^{dec}$, which will be processed by the  multi-exit task-specific heads to output early-exit results.

\subsubsection{Multi-exit task-specific heads}
The different tasks require different numbers of layers due to the reasons below: (i) Each task is of different difficulty and requires different numbers of layers to achieve satisfactory performance, and the multi-exit architecture can provide early-exit results for simple tasks; (ii) Different tasks require different levels of semantic information from various layers. As shown in Fig. \ref{Multi-exit}, we attach the task-specific heads to these intermediate layers. Then the inference time of some simple tasks can be significantly reduced. The difficulty of the task can be determined by the size of the dataset. As for the aforementioned five tasks, the VQA is the most difficult task and requires the maximum number of layers, while the sentiment analysis is the easiest one that exits the earliest.  

\subsection{Channel Encoder and Decoder} 
The encoded features are compressed by the channel encoder and decompressed by the channel decoder as well as eliminating the signal distortion caused by the wireless channel. The channel encoders and decoder are modeled as the the fully-connected networks with ReLU as the activation function.

\subsection{Training Method}
To jointly learn five tasks, we propose an efficient method to train the modules in the U-DeepSC system with domain adaptation. We will randomly take samples from two tasks at one time, and the task with larger dataset will be assigned a higher sampling probability, since they are generally more difficult to learn. Besides, the domain adaptation loss is adopted in the training procedure to specify the task-specific features. It is added by the values of loss function computed on the current two tasks to perform stochastic gradient descent together for learning two tasks simultaneously. The detailed training procedures are summarized in Algorithm \ref{Training}.

\begin{algorithm}[t] 
	\begin{small}
		\caption{Training procedures of the U-DeepSC} 
		\label{Training}
		\DontPrintSemicolon
		\SetKwInOut{Input}{Input}
		\SetKwInOut{Output}{Output}
		\SetKwInOut{Initialize}{Initialize}
		\Input{Five training datasets consist of input images and texts with their labels, and batch size $B$.}
		\Output{The trained U-DeepSC model with encoders and decoders.}
		\For{$m\leftarrow 1$ \KwTo $M$}{
			Randomly choose two tasks and generate two mini-batch samples, A and B from the corresponding datasets. \\
			Compute the task-specific loss $\mathcal{L}_{A}$ and $\mathcal{L}_{B}$ based on the loss function of each task.\\
			Compute the domain adaptation loss $\mathcal{L}_{a}$ based on (\ref{ada}).\\
			Compute the total loss $\mathcal{L}=\mathcal{L}_{A}+\mathcal{L}_{B}+\mathcal{L}_{a}$.\\
			Train the model based on $\mathcal{L}$.
		}
	\end{small}
\end{algorithm}

\section{Simulation Results} \label{Simulation}

In this section, we test our U-DeepSC on the aforementioned five tasks, and five datasets are considered. In particular, Cars196 and CIFAR-10 datasets are adopted for image retrieval and image reconstruction tasks, respectively. As for the text reconstruction and sentiment analysis, the proceedings of the European Parliament and the SST-2 datasets are used, respectively. Moreover, the VQAv2 dataset is used for the VQA task. The image semantic encoder of U-DeepSC is built with eight Transformer encoder layers, and the text semantic encoder is designed with eight Transformer encoder layers. The unified semantic decoder consists of eight Transformer decoder layers. The setting of the training procedure is the AdamW optimizer with learning rate $1\times10^{-4}$, batch size $32$,  weight decay $5\times10^{-3}$.  According to the experimental results of these tasks on different layers, we choose the reasonable layer numbers for these tasks, and the numbers of layers for the VQA, image retrieval, image, reconstruction, text reconstruction, and sentiment analysis are set as $8$, $6$, $4$, $3$, $2$, respectively.

For comparison, three benchmarks are considered. 
	\begin{itemize}
	\item Conventional methods: The conventional separate source-channel coding. For the image data, the joint photographic experts group (JPEG) and the low-density parity-check code (LDPC) are adopted as image source coding and image channel coding, respectively. For the text data, the 8-bit unicode transformation format  (UTF-8) encoding and the Turbo coding are adopted as the text source coding and text channel coding, respectively. The coding rate of channel coding is set as 1/2.

	\item T-DeepSC: The task-oriented deep learning enabled semantic communication (T-DeepSC) designed for a specific task with the same architecture as U-DeepSC, and is implemented by separately trained U-DeepSC.
	
	\item Upper bound: Results obtained via delivering noiseless image and text features to the receiver based on the T-DeepSC.
\end{itemize}

Fig. \ref{results} illustrates the performance of the investigated schemes versus the SNR for different tasks. The proposed U-DeepSC is trained with SNR = $0$ dB and tested in SNR from $-6$ dB to $18$ dB. It is readily seen that both the U-DeepSC and the T-DeepSC outperform the conventional schemes. The U-DeepSC approachs the upper bound at high SNR. It is readily seen that the proposed U-DeepSC achieves approaching performance to the T-DeepSC the in all considered tasks. It shows that our proposed U-DeepSC is able to simultaneously handle 5 tasks with comparable performance to the task-oriented models designed for a specific task. Since only specifc part of the overall features are transmitted in U-DeepSC for different taks, the satisfactory performance of U-DeepSC shows that the task-specific semantic information can be well segmented and specified by the U-DeepSC.

\begin{table}[!hbt]
	\caption{Performance of U-DeepSC trained with or without domain adaptation loss.}
	\renewcommand{\arraystretch}{1.4}
	\centering
	\begin{tabular}{|c|c|c|c|c|}
		\hline
		Task    & \makecell{without domain \\ adaptation Loss}       & \makecell{with domain\\ adaptation Loss}  \\ \hline
		Image Retrieval   & 70.0  & 73.9  \\ \hline
		VQA                          & 57.8   & 60.9    \\ \hline
		Text Reconstruction     &0.94    & 0.96\    \\ \hline
		Image Reconstruction        & 31.9   & 32.0  \\ \hline
		Sentiment Analysis         & 80.1    & 84.2   \\ \hline
	\end{tabular}
	\label{table1}
\end{table}

Table \ref{table1} depicts the performance of U-DeepSC trained with or without domain adaptation loss. It has been proved that the domain adaptation loss significantly improves the performances of image retrieval, VQA, and sentiment analysis. However, the performances of the text reconstruction and image reconstruction are almost unchanged, which is mainly because all of the encoded features are transmitted for these two tasks, i.e., the semantic information of the overall encoded features will keep the same in both cases.

\begin{table}[!hbt]
	\caption{Number of parameters.}
	\renewcommand{\arraystretch}{1.3}
	\centering
	\begin{tabular}{|c|c|c|c|c|}
		\hline
		Task    & T-DeepSC        & U-DeepSC   \\ \hline
		Image Retrieval   & 46.9M   & 95.6M  \\ \hline
		VQA                          & 92.3M   & 95.6M    \\ \hline
		Text Reconstruction     &52.9M    & 95.6M\    \\ \hline
		Image Reconstruction        & 42.5M   & 95.6M  \\ \hline
		Sentiment Analysis         & 55.6M    & 95.6M   \\ \hline
		Stored Parameters         & 290.2M    & 95.6M   \\ \hline
	\end{tabular}
	\label{table2}
\end{table}
 As shown in Table \ref{table2}, the total stored number of parameters of T-DeepSC is 290.2M, which is obtained by adding the parameters required for each task. For our proposed U-DeepSC, the number of stored model parameters is only 95.6M for five tasks, which is 67.1\% less than that of the T-DeepSC. The U-DeepSC is able to provide satisfactory performance with much-reduced model parameters. It is of great significance towards a practical semantic communication system for scenarios with limited spectrum resources and storage resources.

\section{Conclusion} \label{Conclusion}
In this paper, we firstly proposed a general framework for U-DeepSC. Particularly, we considered five popular tasks and jointly trained these tasks with a unified model. To learn the unified model for serving various tasks and achieve task-specific transmission overhead, we employed domain adaptation in the training procedure to specify the task-specific features for each task. Thus, only the task-specific features were transmitted in U-DeepSC. Then, we developed a multi-exit architecture to provide early-exit  results for relatively simple tasks, which reduced the inference time.  Simulation results showed that our proposed model has satisfactory performance in low SNR regime, and achieved comparable performance to the task-oriented model designed for a specific task.

\bibliographystyle{IEEEtran}
\bibliography{IEEEabrv,Semantic}

\end{document}